# Topological Descendants of Multicritical EuTl$_2$

Lin-Lin Wang[1], Hoi Chun Po[2], Robert-Jan Slager[3,4], and Ashvin Vishwanath[4]

[1]Ames Laboratory, U.S. Department of Energy, Ames, IA 50011, USA
[2]Department of Physics, Massachusetts Institute of Technology, Cambridge, MA 02139, USA
[3]TCM Group, Cavendish Laboratory, University of Cambridge, J. J. Thomson Avenue, Cambridge CB3 0HE, United Kingdom
[4]Department of Physics, Harvard University, Cambridge, MA 02138, USA

## Abstract

The past decades have witnessed a transformation in characterizing condensed matter systems with topology. Aided by a refined understanding of topological band structures with crystalline symmetries that has emerged recently, many electronic phases have been identified and a plethora of materials have been predicted to host novel properties and functionalities. A key underlying question, also with respect to future application, is to what extent the related physical features can be manipulated, especially in the context of magnetic order. Here we describe a paradigmatic semimetal that simultaneously incorporates multiple, and sometimes conflicting topology which guarantees gaplessness and leads to an exceptionally rich family of descendent phases on lowering symmetry. We predict that this multicritical phase is realized in EuTl$_2$. Starting from the parent semimetallic state, which already separates two topological insulating regimes, the interplay of inherent magnetism and strain allows for an exceptionally rich phase diagram of topological descendant states.

# Introduction

An important outstanding question in the field of topological materials[1-4] is how their robust physical properties, such as anomalous surface states and quantized responses, could be leveraged for technological applications. Combining nontrivial electronic topology[5-13] with magnetism[14] provides a promising route forward, with the prospect that the presence or absence of certain robust responses could be tuned dynamically through an external magnetic field. Such sensitivity of topological character on magnetism, however, is not always guaranteed in a magnetic topological material. For example, a magnetic axion insulator protected by inversion symmetry[15-17] could be viewed as the natural continuation of a parent paramagnetic or nonmagnetic strong topological insulator (TI), in the sense that no bulk band gap closing is required in going between the two phases. This is the case in the recently discovered anti-ferromagnetic (AFM) TI in the $MnBi_2Te_4(Bi_2Te_3)_m$ series[18-24] and the predicted AFM higher-order TI and topological crystalline insulator (TCI) in $EuIn_2As_2$[25].

A natural paradigm for discovering materials with the desired strong interplay between magnetism and electronic topology is to focus on semimetallic parent phases with time-reversal symmetry (TRS) which are symmetry-tuned to a multicritical point between different topological phases. If such a phase is realized in a compound with an innate magnetic instability, then different topological phases could be naturally spawned in conjunction with different magnetic orders. In this work, we theoretically propose $EuTl_2$ as a prototype candidate for this paradigm. In the parent phase, the system is a Dirac semimetal (DSM) whose gaplessness could be attributed to the competition between the opposing tendency to realizing a strong TI and a TCI. A plethora of distinct topological phases descends from this parent phase upon the breaking of symmetries either by strain or by the development of magnetic orders as summarized in Fig.1. In particular, the set of daughter phases include (i) paramagnetic TCIs protected by different spatial symmetries with two sets of mirror Chern numbers (MCN), (ii) AFM Dirac semimetal (AFM-DSM), (iii) AFM axion insulator, and (iv) ferromagnetic (FM) Weyl semimetal (FM-WSM). The set of topological phases realizable in this system are summarized in Table 1 together with their associated space groups, magnetic space groups and the relevant topological indices.

# Results and Discussion

**Magnetic topological phases from a multicritical parent**

We begin by elaborating on how a multicritical parent phase with TRS could serve as a promising starting point for the search for topological magnetic phases. Topologically nontrivial insulators come in two varieties: Chern insulators like that realized in the Haldane model[26] and anomalous quantum Hall effect[27] do not require symmetry protection, whereas phases like the quantum spin-Hall insulator and TCIs require protecting symmetries, e.g. TRS and crystalline symmetries. In some systems the nontrivial ground state could be simultaneously protected by multiple symmetries. A common example concerns centrosymmetric three-dimensional strong TI with significant spin-orbit coupling (SOC) and TRS. The nontrivial magnetoelectric response is quantized provided at least one of spatial inversion and time-reversal symmetries is preserved. As such, if inversion-preserving magnetism is induced in the strong TI, one naturally arrives at a magnetic TI which is dubbed an "axion insulator." This design strategy has led to the discovery of magnetic TIs in the $MnBi_2Te_4(Bi_2Te_3)_m$ series[18-24].

A corresponding dichotomy on the stability of topological semimetal (TSM) with respect to symmetry breaking could also be advanced for TSMs: while 3D WSMs are stable for a finite range of system parameters, the 3D DSM requires symmetry protection, and as such could be gapped out by an infinitesimal symmetry-breaking perturbation. However, there is an important distinction in the semimetallic case, since generically distinct gapped phases could be realized by the symmetry lowering. Equivalently, symmetry-protected TSMs could be viewed as being pinned to a critical point between topologically distinct phases. Indeed, it is known that the 3D DSMs can be viewed as a critical point between phases with different strong and weak $Z_2$ indices, and these phases could be accessed by applying suitable strains to the crystal[28]. The TSM could even be multicritical, in the sense that more than two descendent topological phases could be realized with different symmetry breaking patterns. This is especially true in the presence of crystalline symmetries, which leads to a rich set of topological phases mirroring the corresponding richness of spatial symmetry patterns in crystals[9]. Consequently, TSMs inherently have a richer phase diagram as a function of symmetry-lowering perturbations. That is, they could remain stable, be converted into a different variety of TSM (e.g. splitting a Dirac point to Weyl points), or become distinct gapped topological phases depending on the sign of the perturbation. As a concrete example, let us turn to the phases realizable in EuTl$_2$, which is a DSM protected by the 3-fold rotation symmetry ($C_3$) in the paramagnetic phase. The spatial symmetries include two sets of mirror planes which are respectively horizontal or vertical, and as such we could define MCNs for each of the planes. However, the Dirac points lie on the $C_3$ rotation axis and render the vertical MCN ill-defined. Correspondingly, upon lattice distortion with strain to remove $C_3$ and gap out the Dirac points, the horizontal MCN remains unchanged but the vertical MCN becomes either 2 or 0, depending on the sign (i.e., tensile vs. compressive) of the strain. This shows explicitly how changing the sign of the strain could result in different TCI phases.

There are, however, three main drawbacks in using strain as an experimental knob to access the phase diagram around a symmetry-protected TSMs: first, the crystal symmetries may conspire to leave only a single natural straining direction, making it impossible to access some of the descendant phases of the parent multicritical TSM. Second, tensile strain is generally harder to sustain than compressive strain in a crystal, which means it may not be physically feasible to access both sides of the phase diagram. Finally, the attainable strain level in a crystal is usually small (at a few %), and, unless the system is close to a structural instability, this translates into a quantitatively small effect on the electronic properties.

The incorporation of magnetism into the problem provides an important new route, as its richness and spontaneous nature provide access to a much larger phase space starting from a multicritical parent TSM. In addition, for systems with competing magnetic states it may be possible to select different magnetic orders by changing growth conditions, as recently shown for EuCd$_2$As$_2$ in experiment, where both AFM and FM samples[29] can be grown. The according control is one of the most important outstanding challenges in converting electronic topology into topological electronics. As such, searching for a multicritical TSM parent with magnetic descendant topological phases could be a useful guiding principle in functionalizing topological materials. For EuTl$_2$ (Fig.1), different magnetic configurations can give a range of different topological phases, which are also summarized in Table 1. Below we will discuss these topological phases in EuTl$_2$ with details, and show that the multicitical behavior indeed results in a myriad of phases. Such abundance of descendent topological phases stands in rather stark contrast with the stable gapped AFM order in, e.g., the MnBi$_2$Te$_4$(Bi$_2$Te$_3$)$_m$ [18-24].

**EuTl$_2$ as a case example of a multicritcal parent**

The crystal structure of EuTl$_2$ (Fig.2(a)) in space group 194 (P6$_3$/mmc) has two oppositely buckled honeycomb Tl layers and both are sandwiched between two hexagonal Eu layers. The space group symmetry generators are the inversion $P$, 3-fold rotation $C_3$, 2-fold screw $S_2$ along the $c$-axis, and 2-fold rotation along the $a$ or $b$-axis, labelled as $C_{2;[010]}$ or $C_{2y}$. With the Cartesian coordinate system chosen for $y$ ($z$) along $b$ ($c$)-axis, the Miller index of $[1\bar{2}10]$ and $[10\bar{1}0]$ in hexagonal coordinate is equivalent to [010] and [100] in Cartesian coordinates, respectively. The 6-fold screw along the $c$-axis, $S_6$, is composed of $C_3$ and $S_2$. There is a mirror symmetry, labeled as $M_{(001)}$ or $M_z$, about the horizontal plane normal to the $c$-axis. Similarly, the symmetries about another set of vertical mirror planes normal to the $a$ or $b$-axis are labeled as $M_{(010)}$ or $M_y$. There are also vertical glide planes $G_{(100)}$ or $G_x$, spanned by $c$-axis and $a$ or $b$-axis, and we denote the 2-fold rotation about the normal of the glide plane by $C_{2;[100]}$ or $C_{2x}$. The inversion center locates at the midpoint of the Tl-Tl bond in the buckled honeycomb layer. Figure 2(b) shows the 3D bulk Brillouin zone (BZ) of EuTl$_2$ and the 2D surface Brillouin zone (SBZ) for the (001) and (010) surfaces with high-symmetry points labeled.

The bulk band structures calculated with PBE+SOC is plotted in Fig.2(c) for nonmagnetic EuTl$_2$, where the half-filled Eu 4$f$ orbitals are treated as core states. There is a continuous gap between the top valence and bottom conduction bands except for the crossing point along the $\Gamma$-$A$ direction. The band dispersion near the crossing point, zoomed in Fig.2(d), shows that the gap vanishes in conjunction with a change in orbital projection of Tl $p_y$. Note that there is a near degeneracy along the $\Gamma$-$K$ line but a small gap is sustained. As such, we conclude the most important degrees of freedom near the Fermi energy (E$_F$) is the pair of bulk Dirac points (BDPs) at momentum-energy (0.0, 0.0, ±0.36 Å$^{-1}$; E$_F$–0.19 eV). On one hand, the BDPs are protected by $C_3$, since the states crossing at the BDP have distinct $C_3$ eigenvalues. On the other hand, the parity eigenvalues at time-reversal invariant momentum (TRIM) gives a nontrivial $Z_4$-valued symmetry indicator of 2, indicating the system could be viewed as a would-be TCI if not for the presence of the BDPs. Because the gaplessness is only along the $\Gamma$-$A$ direction, the $M_z$ MCN[30, 31] on the $k_z$=0 and $\pi$ planes are well-defined and calculated as 2 and 0, respectively, from the Wannier charge center (WCC) evolution (see Fig.S1 in SI).

We now address how the topological features in the bulk manifest on the surfaces. From the surface states on (001) in Fig.2(e), there are surface Dirac points (SDPs) at the three TRIM points, $\bar{M}$. Such SDPs should be protected by $M_y$ symmetry. But the $M_y$ MCN cannot be calculated because of the gaplessness along the $\Gamma$-$A$ direction on the $k_y$=0 plane. As zoomed along $\bar{K}$-$\bar{\Gamma}$-$\bar{M}$ direction in Fig.2(g), surface states converging to the BDP projection on (001). Together with the three SDPs at $\bar{M}$, there are four gapless points on (001). In Fig.2(f), we also show the surface state spectrum on the (010). As the glide symmetry is preserved, the surface states along the $\bar{X}$-$\bar{S}$-$\bar{Y}$ direction correspond to that of the hourglass[32] fermions. The double degeneracy of surface state along $\bar{S}$-$\bar{Y}$ is protected by $TG_x$ and the hourglass crossing (zoomed in Fig.2(h)) is protected by $G_x$. All in all, we find that the bulk and surface band structures of EuTl$_2$ showcase three types of topological features protected by different symmetries, namely, bulk Dirac point, surface Dirac points and hourglass fermions.

**Descendent states under strain**

As a next step, we consider the topological phases descending from the high-symmetry parent phase characterized above. We will first consider in details the effects of a uniaxial strain along the hexagonal [1$\bar{1}$00] direction, which breaks the $C_3$ symmetry to create an insulating phase by gapping out the BDP along the $\Gamma$-$A$ direction, while preserving every other symmetry element (see Table 1). The space group is changed from 194 to 63. This analysis will then serve as the foundation upon which we can readily understand the topological characters of the various magnetic phases. As shown in Fig.3(a), with a –2% strain, a 6 meV gap is opened at the innate BDP. Correspondingly, on the (001) a gap opens up in the projection of the bulk bands and a SDP emerges (Fig. 3(b)). As the SDPs at the other three TRIM are unaffected by the perturbation (Fig.3(d)), we see that the system has an even number of SDPs and is therefore not a strong TI, in distinct contrast to BaHgSn[33] and KAuTe[34]. This still leaves the possibility that EuTl$_2$ under strain is a TCI. To investigate this, we compute the $M_y$ MCN, which is now well-defined with the BDP being gapped out. The calculated $M_y$ MCN is 2 and 0 for –2% and +2% strain, respectively, as shown in Fig.3(c) and (e). Together with the unchanged $M_z$ MCN of 2, this certifies that the system is a TCI at least for compressive strain. Furthermore, if one views the undistorted phase as a critical point between the two classes of strained structures, then it would be natural to assign $M_y$ MCN=1 to the high-symmetry parent phase despite the BDPs render the $M_y$ MCN ill-defined in a strict sense. This picture helps to elucidate on the gaplessness of the system: on the one hand, from considerations of both the symmetry indicators and the $M_z$ MCNs, we see that the system is not a strong TI; on the other hand, the natural $M_y$ MCN assignment to the undistorted phase is only consistent with a strong TI. The presence of the BDPs could then be viewed as the resolution between the contradictions derived from considering different sets of spatial symmetries.

We characterize the nature of the TCI realized in EuTl$_2$ with tensile strain. The strain does not break the $M_z$ or $G_x$ symmetries, and the small perturbation has little effect on the Bloch states on the $k_z$=0 and $\pi$ planes. As such, the $M_z$ MCN of 2 and 0 on these planes for the higher-symmetry structure survives for both compressive and tensile strain. In fact, the topological characters of an insulator with respect to multiple crystalline symmetries are intertwined[9, 10, 35, 36], and based on the MCNs for $M_y$ and $M_z$ we have computed, one can infer the TCI indices associated with the other crystalline symmetries. The results are summarized in Table 1, and we see that EuTl$_2$ realizes distinct composite TCI phases depending on the nature of the strain.

Inspecting the surface state signatures, however, reveals a conundrum. As one can see in Fig.3(f) and (d), energetically the surface band structures on the (001), which preserves $M_y$, are almost identical for the tensile and compressive strains despite the system is respectively trivial and nontrivial as a $M_y$-protected TCI. Such surface behavior is nevertheless consistent with the bulk diagnosis: on a mirror-respecting surface, one could define a Z-valued topological invariant to each of the SDPs[30], and the defining feature of the anomalous surface of a nontrivial TCI bulk is that the total charge (even number) of SDPs is nonzero. Since the SDPs are all pinned to the TRIM points, the topological distinction between the surface states with opposite strain cannot be discerned from the energetics alone, i.e., simply noting the distribution of the SDPs in the surface Brillouin zone does not allow one to uniquely determine the topological index of the system. Instead, one has to analyze the system from the bulk perspective, as we have done here, or evaluate the mirror-protected Z-valued topological invariants for each of the SDPs[30].

**Descendent states with magnetism**

Next, we discuss the topological phases that emerge when magnetism enters, which is natural in the system given that the Eu half-filled 4$f$ orbital provide a large local magnetic moment[25, 29, 37-39]. So far there has been no experimental studies on the magnetic ground state of EuTl$_2$. In this work we consider the descendent phases originating from the development of commensurate A-type AFM orders (AFMA), i.e. ferromagnetic coupling within the Eu layer on the *ab*-plane and anti-ferromagnetic coupling along the stacking *c*-axis, as well as the FM order along the *c*-axis. From our DFT total energy calculations (see Table S1 in SI), although the FM order is preferred with the experimental structural parameters, after full relaxation in PBE+U+SOC, the AFM orders become more stable. For hexagonal Eu compounds, in contrast to the earlier reports of AFMA[40,41], recent experiment on EuCd$_2$As$_2$ found that the magnetic order can be manipulated by changing growth conditions to get either AFM or FM samples[29]. We caution that incommensurate orders, like the helical order observed in EuIn$_2$As$_2$[42], are also possible. The interplay between such incommensurate magnetic orders and the descendent topological phases from the multicritical parent is an interesting open question, but it is out of the scope of the present work.

For AFMA configuration, both time-reversal (*T*) and inversion (*P*) are broken. Additionally, for AFMA with moment along *c*-axis (AFMA*c*), the $S_2$ and $M_y$ are broken, but the $C_3$, $M_z$ and $G_x$ are retained. In contrast, for AFMA with moment along *b*-axis (AFMA*b*), the $C_3$ and $M_z$ are removed, but the $S_2$, $M_y$ and $G_x$ are retained. These changes of symmetry elements for different magnetic configurations will have consequences for the band structure and topological properties. We will show that the different topological characters are selectively tuned in the different magnetic configurations based on our understanding of the composite topological features in the parent TRS-protected EuTl$_2$.

***AFMAc Dirac semimetal***: Although $T$ and $P$ are both broken in the AFMA configurations, the ($TP$) combination can still give Kramer pairs and band double degeneracy due to the anti-unitary $(TP)^2 = -1$. The bulk band structures of AFMA$c$ in Fig.4(a) is similar to the nonmagnetic case with a BDP along $\Gamma$-$A$ direction protected by $C_3$, as shown in Fig.4(b). One difference is that the extra band degeneracy along the $A$-$L$ direction on $k_z=\pi$ plane in the nonmagnetic is lifted in AFMA$c$ for breaking the non-symmorphic $S_2$. Turning to the surface state signature for this pair of AFM-BDP, there are four surface bands converging to the BDP projection on the (001) in Fig.4(e). The corresponding Fermi arcs are clearly shown in Fig.4(f) on the (100). In contrast, the SDPs at the three $\bar{M}$ on (001) are all gapped out due to the breaking of both $T$ and $M_y$, as shown in Fig.4(c). In contrast, $M_z$ is still a good symmetry and the calculated $M_z$ MCN changes from 2 to 1 on $k_z=0$ plane, while it remains as 0 on $k_z=\pi$ plane (see Fig.S3 in SI), implying the system could become an axion insulator if the BDPs could be gapped out by a small perturbation. On the (010) surface, there are still hourglass features, but the crossing is pushed from $\bar{X}$-$\bar{S}$ to $\bar{S}$-$\bar{Y}$ direction as seen in Fig.4(d). Comparing to the parent EuTl$_2$, the SDP features protected by $M_y$ is selectively removed in AFMA$c$.

Around the BDP, the band inversion is between the bonding Tl $p_x$-$p_y$ band and anti-bonding Tl $p_z$ band with some Eu $s$ character. For PM in space group (SG) 194, the irreducible representations are $\Gamma_7^+$ and $\Gamma_9^+$ for the two doubly degenerated bands, respectively, and a low-energy effective 4-band Hamiltonian similarly to BaAuBi[43] and Na$_3$Bi[44] can be constructed under $T$-invariant. In contrast, although the pair of BDPs in AFMA$c$ is still protected by $C_3$, the combined symmetry ($TP$) means the Hamiltonian for AFMA$c$ should not be $T$-invariant anymore. For AFMA$c$, the magnetic space group (MSG) is 194.266. For the co-representations of this type-III MSG, the unitary SG is 190. We have used theory of invariant[45] to construct the 4-band $k \cdot p$ model by considering the Dirac $\Gamma$ matrices and polynomials that are compatible under the symmetry transformations in MSG-194.266.

$$H(\boldsymbol{k}) = \varepsilon_0(\boldsymbol{k})\mathbb{I}_{4\times4} + M(\boldsymbol{k})\tau_z\sigma_0 + H_1(\boldsymbol{k}) + H_2(\boldsymbol{k}) \qquad (1)$$

$$H_1(\boldsymbol{k}) = Bk_z\tau_y\sigma_0 + A(k_x\tau_x\sigma_y + k_y\tau_x\sigma_x) \qquad (2)$$

$$H_2(\boldsymbol{k}) = D\left((k_x^2 - k_y^2)\tau_x\sigma_y + 2k_xk_y\tau_x\sigma_x\right) \qquad (3)$$

where $\sigma$ and $\tau$ are Pauli matrices for spin and orbital, respectively.

Among the different terms in $H(\mathbf{k})$, $H_2(\mathbf{k})$ breaks $T$, while the whole Hamiltonian is invariant under the combined symmetry ($TP$). After unitary transformation to make the coefficients of the matrix elements real, the Hamiltonian can be written as,

$$H(\mathbf{k}) = \varepsilon_0(\mathbf{k}) + \begin{pmatrix} M(\mathbf{k}) & Bk_z & 0 & Ak_+ + Dk_+^2 \\ Bk_z & -M(\mathbf{k}) & -Ak_+ - Dk_+^2 & 0 \\ 0 & -Ak_- - Dk_-^2 & M(\mathbf{k}) & Bk_z \\ Ak_- + Dk_-^2 & 0 & Bk_z & -M(\mathbf{k}) \end{pmatrix} \quad (4)$$

where $k_\pm = k_x \pm ik_y$, $M(\mathbf{k}) = M_0 - M_1 k_z^2 - M_2(k_x^2 + k_y^2)$, and $\varepsilon_0(\mathbf{k}) = C_0 + C_1 k_z^2 + C_2(k_x^2 + k_y^2)$. After diagonalization, the energy eigenvalues are

$$E(k) = \varepsilon_0(\mathbf{k}) \pm \sqrt{M(\mathbf{k})^2 + B^2 k_z^2 + A^2(k_x^2 + k_y^2) + 2ADk_x(k_x^2 + k_y^2) + D^2(k_x^2 + k_y^2)^2} \quad (5)$$

When fitted to DFT calculated band structures, the following parameters are obtained, $C_0 = -0.081$ eV, $C_1 = -1.186$ eVÅ², $C_2 = 42.777$ eVÅ², $M_0 = 0.736$ eV, $M_1 = 5.429$ eVÅ², $M_2 = 32.573$ eVÅ², $A = 1.463$ eVÅ, $B = 0.000$ eVÅ and $D = 0.086$ eVÅ². The BDP in AFMA$c$ has the momentum-energy of (0.0, 0.0, ±0.37 Å$^{-1}$; $E_F$–0.25 eV).

*AFMAb Axion insulator*: When the AFMA configuration has magnetic moments in-plane along $b$-axis (AFMA$b$) (Fig.5(a)), the $C_3$ and $M_z$ are broken, while $S_2$, $M_y$ and $G_x$ survive. The system still has the ($TP$) symmetry and the band double degeneracy. The band structure of AFMA$b$ in Fig.5(a) shows that the BDP is gapped out due to the breaking of $C_3$. In contrast to the gap due to strain resulting in a TCI with $M_y$ MCN of 2 or 0, the WCC evolution in Fig.5(b) shows that $M_y$ MCN is 1, thus the system becomes an axion insulator. This should be contrasted with the nonmagnetic strain-induced gapping, since in the that case $T$ dictates that the MCN contribution coming from each of the BDPs is the same, and so the resulting MCNs are 1±1 = 0 or 2. In contrast, the MCN contribution from the two BDPs cancel each other when going into the AFMA$b$ phase, revealing the underlying $M_y$ MCN=1, which one could have attempted to assign to the high-symmetry phase. As seen from Fig.5(c), only a single SDP survives in the vicinity of the $\bar{M}$ point protected by $M_y$, among the four gapless SDPs on (001) in nonmagnetic EuTl$_2$. There is no SDP at the $\bar{\Gamma}$ point (see Fig.S4 in SI). On the (010) surface, the hourglass features in Fig.5(d) also give an odd number of surface band crossings, as is expected for an axion insulator. Another notable feature is that the original Dirac point at the $\bar{X}$ point is now moved slightly away towards to $\bar{S}$ point.

*FMc Weyl semimetal*: Lastly, we consider the band structure of the FM configuration with moments along the *c*-axis (FM*c*). For FM*c*, *T* is broken, but *P* and the $S_2$ remain. There is no combined (*TP*) symmetry. As shown in Fig.6(a), with breaking *T* but preserving *P*, the band double degeneracy is all lifted except for along the high-symmetric *A-L* direction, which is protected by the $S_2$. There are 14 pairs of Weyl points (WPs), which can be grouped into three sets of symmetry-related pairs, as shown in Fig.6(b). The momentum-energy of all WPs are also listed in Table S2. Set 1 has two pairs of WPs along the *Γ-A* direction by splitting the pair of BDP. Set 2 of six pairs of WPs are parallelly separated away from the $k_z=0$ plane, while the Set 3 of another six pairs of WPs are along the *Γ-K* direction on the $k_z=0$ plane. The Berry curvatures of the selected pairs of WPs in each set have been plotted in Fig.S5 to confirm their monopole features (see Fig.S5 in SI). For Set 1 with the two pairs of WPs split from the pair of BDP, one pair is at $E_F$–0.18 eV and the other at $E_F$–0.24 eV. When projected on (001), two sets of surface bands converging to the two sets of WPs separately at different energies in Fig.6(c), which can be compared to Fig.4(e) with all four surface bands converging to the same projected point for the BDP pair. Similarly, for the Fermi arcs in Fig.6(d) coming out from the WP projections on (100), comparing to the closed Fermi arcs in Fig.4(f) for the BDP pair, the arcs are split into two open pieces, giving the surface signature of WPs.

## Conclusions

In conclusion, we have shown that EuTl$_2$ is a paradigmatic multicritical topological material enabling the access of a plethora of distinct topological phases by tuning strain and magnetic configurations. Remarkably, the phases realized include both Dirac and Weyl semimetals, the axion insulator, and a variety of other topological crystalline insulators protected by the combinations of different spatial symmetries. Correspondingly, the associated surface state signatures include surface Dirac points which may or may not be pinned to high-symmetry points, as well as hourglass fermions. Our calculations show that the relation between the surface dispersion and the bulk topological diagnosis could be rather subtle, and therefore caution must be used in viewing the surface states as a smoking-gun signal of the topological phase realized. The prospect of using strain and magnetic field to drive phase transitions between different topological phases warrants further experimental investigations.

## Acknowledgements

We thank Paul C. Canfield and Adam Kaminski for inspiring discussions. This work was supported as part of the Center for the Advancement of Topological Semimetals, an Energy Frontier Research Center funded by the U.S. Department of Energy Office of Science, Office of Basic Energy Sciences through the Ames Laboratory under its Contract No. DE-AC02-07CH11358. R.-J. S. in addition acknowledges funding via the Marie Sklodowska-Curie programme [EC Grant No. 842901] and the Winton programme as well as Trinity College at the University of Cambridge. The work of H.C.P. is supported by a Pappalardo Fellowship at MIT and a Croucher Foundation Fellowship.

## Computational Methods

All density functional theory[46, 47] (DFT) calculations with and without spin-orbit coupling (SOC) were performed with the PBE[48] exchange-correlation functional using a plane-wave basis set and projector augmented wave method[49], as implemented in the Vienna Ab-initio Simulation Package (VASP)[50, 51]. Using maximally localized Wannier functions[52, 53], tight-binding models were constructed to reproduce closely the band structure including SOC within $E_F \pm 1eV$ with Eu $s$-$d$-$f$ and Tl $s$-$p$ orbitals. The surface Fermi arcs and spectral functions were calculated with the surface Green's function methods[54, 55] as implemented in WannierTools[56]. In the DFT calculations, we used a kinetic energy cutoff of 250 eV, $\Gamma$-centered Monkhorst-Pack[57] (10×10×6) $k$-point mesh, and a Gaussian smearing of 0.05 eV. To account for the strongly localized Eu $4f$ orbitals, an onsite Hubbard-like[58] U=5.0 eV is used. For band structure calculations of $EuTl_2$, we have used the experimental structural parameters[59], $a = 5.035$ Å, $c = 7.964$ Å with Eu at $2b$ positions of (0, 0, 1/4) and Tl at $4f$ positions of (1/3, 2/3, 0.044).

**Competing Interests**: The authors declare that there are no competing interests.

**Author Contributions**: L.-L.W., H.C.P., R.-J.S. and A. V. conceived and designed the project. L.-L.W. performed ab initio calculations. H.C.P., R.-J.S. and A. V. performed symmetry-based analysis. All authors contributed to the writing and editing of the manuscript.

**Data Availability**: The data that support the findings of this study are available in Materials Data Facility (hyperlink will appear later)

Table 1. Summary of symmetries and topological diagnosis of EuTl$_2$. The paramagnetic high-symmetry phase belongs to space group (SG) 194, and when either tensile or compressive strain is incorporated the spatial symmetries are reduced to those of SG 63. The ordering associated with the magnetic structures AFMA*c*, AFMA*b* and FM*c* are defined in the text, and we indicate their magnetic space groups (MSGs) in the Belov-Neronova-Smirnova notation. Asterisk (*) indicates the bulk is gapless. × indicates the symmetry is broken, and 0 indicates the system does not possess a nontrivial topological index associated with the symmetry. For $M_y$ and $M_z$, we indicate the computed mirror Chern number for the mirror plane containing the $\Gamma$ point. For $P$, we indicate the $Z_4$-valued symmetry indicator appropriate for the symmetry setting (paramagnetic vs. magnetic). For the rest of the symmetries, we indicate the $Z_2$ crystalline invariant one could infer from the invariants associated with $M_y$, $M_z$, and $P$. When the invariant is undetermined due to the gapless nature of the bulk we enter * in the corresponding entry. Note that, for the magnetic phases, a × for an order-two symmetry $G$ (all the symmetries below except for $C_3$) is synonymous with $GT$ being a symmetry, where $T$ denotes time-reversal. In particular, the axion insulator AFMA*b* also automatically realizes the nontrivial phase protected by the $C_2T$ symmetry, and therefore could host hinge modes on suitable edges.

| Structure & Sym. | $C_3$ | $M_y$ | $M_z$ | $P$ | $G_x$ | $C_{2y}$ | $C_{2x}$ | $S_2$ |
|---|---|---|---|---|---|---|---|---|
| Parent* (SG 194) | 0 | * | 2 | 2 | * | * | * | * |
| Tensile strain (SG 63) | × | 0 | 2 | 2 | 0 | 1 | 1 | 0 |
| Compressive strain (SG 63) | × | 2 | 2 | 2 | 1 | 0 | 0 | 0 |
| AFMA*c*  (MSG 194.266) | 0 | × | 1 | × | * | * | × | × |
| AFMA*b* (MSG 63.461) | × | 1 | × | × | 1 | × | × | 0 |
| FM*c** (MSG 194.270) | 0 | × | * | 0 | × | × | × | 0 |

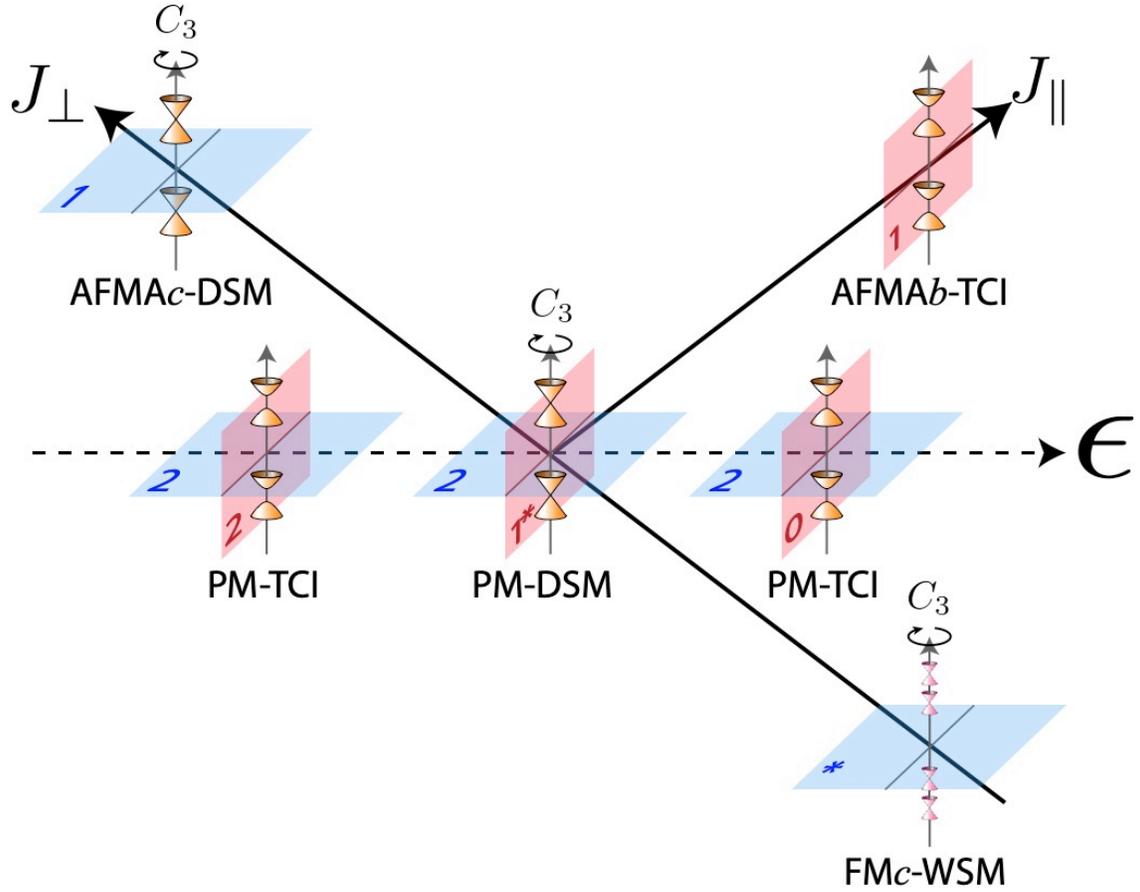

Figure 1. Different topological phases in EuTl$_2$ tuned by strain ($\varepsilon$) and interlayer ($J_\perp$) and intralayer ($J_\parallel$) magnetic couplings. The highest-symmetry paramagnetic (PM) Dirac semimetal in the middle serves as the parent phase and spawns a variety of descendent semi-metallic (SM) or topological crystalline insulator (TCI) phases when the symmetry is lowered. Dirac (D) and Weyl (W) cones are represented by orange and pink cones, respectively. Red and blue shades indicate mirror planes and the associated mirror Chern numbers are engraved. Asterisk (*) indicates the presence of gapless points on the planes, which render the invariants ill-defined. Both A-type antiferromagnetic (AFMA) and ferromagnetic (FM) magnetic structures are considered, with the last character denoting the crystalline axis along which the magnetic moments point. Note that for FM$c$-WSM there are additional Weyl points, some residing on and near the blue mirror plane, which are not shown for simplicity.

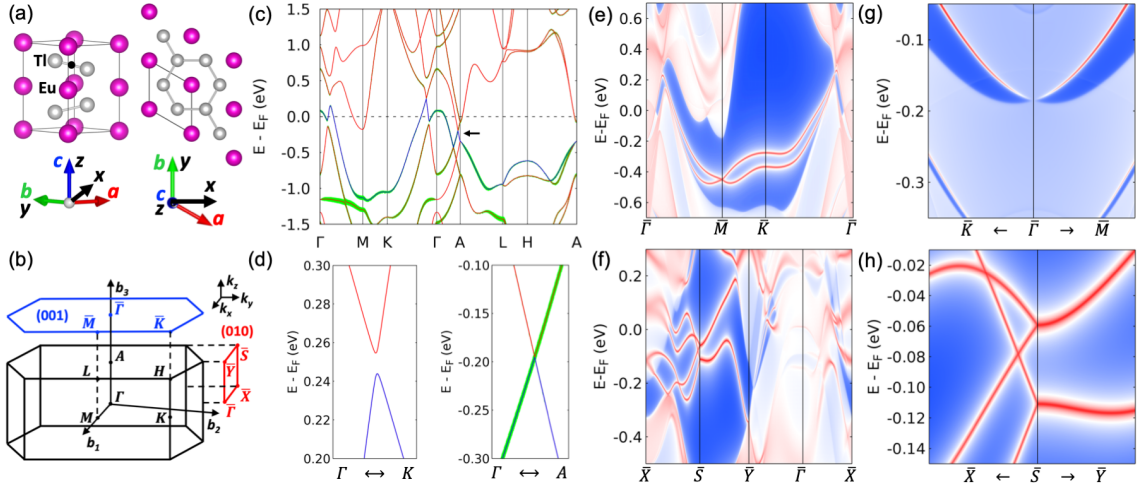

Figure 2. (a) Crystal structure of EuTl$_2$ in space group 194 (P6$_3$/mmc) in both side view and top view. Eu (Tl) atoms are magenta (gray) spheres. The inversion center at mid-point of Tl-Tl bond in the buckled honeycomb layer is marked by a black circle. (b) 3D bulk Brillouin zone (BZ) of EuTl$_2$ and 2D surface Brillouin zone (SBZ) on (001) and (010) with high-symmetry points labeled. (c) Bulk band structure of nonmagnetic EuTl$_2$ in PBE+SOC without 4$f$ orbitals and (d) zoomed along $\Gamma$-$K$ and $\Gamma$-$A$ directions. The top valence band is in blue and the green shade stands for the projection of Tl $p_y$ orbitals. Black arrow points to the bulk Dirac points along $\Gamma$-$A$. Panel (e) and (f) are surface states of (001) and (010) with features zoomed in (g) and (h), respectively. Blue, white and red colors stand for low, medium, and high density of states.

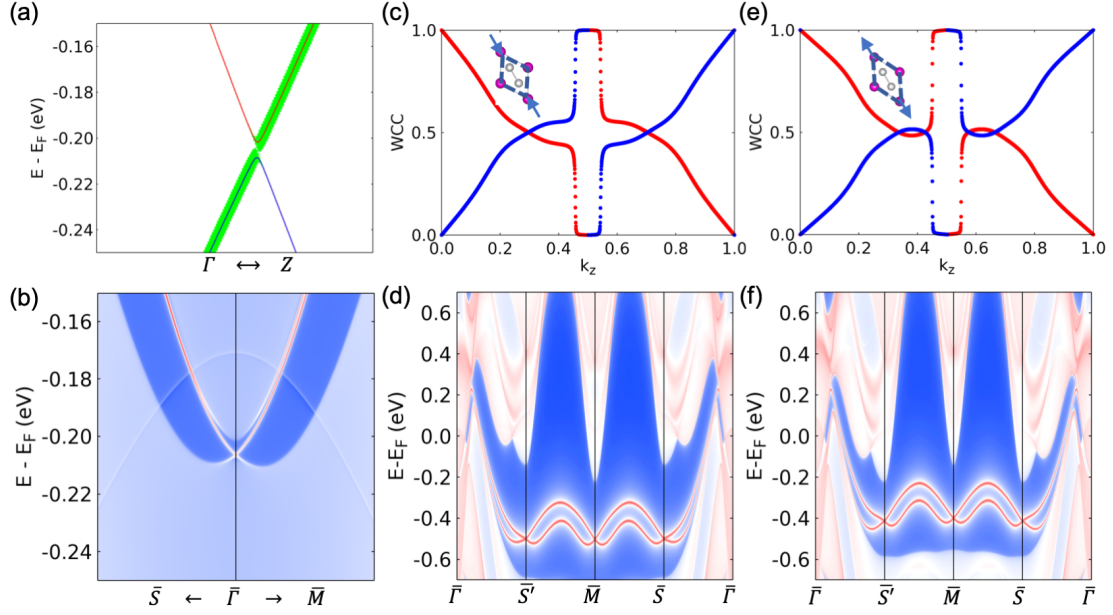

Figure 3. (a) Bulk band structure of nonmagnetic EuTl$_2$ with −2% strain zoomed along $\Gamma$-$Z$ direction in the base-centered orthorhombic structure (space group 63) showing the bulk gap opening. The valence band is in blue and the green shade stands for the projection of Tl $p_y$ orbitals. (b) (001) surface states zoomed along $\bar{S}$-$\bar{\Gamma}$-$\bar{M}$ direction showing the surface Dirac point at $\bar{\Gamma}$ wtih −2% strain. Panel (c) and (e) are the Wannier charge center (WCC) evolution for calculating the $M_y$ mirror Chern number (MCN) of 2 and 0 at −2% and +2% strain, respectively. Panel (d) and (f) are the (001) surface states at −2% and +2% strain, respectively, in the base-centered orthorhombic structure. Note that the surface band structures are essentially insensitive to the different bulk topologies.

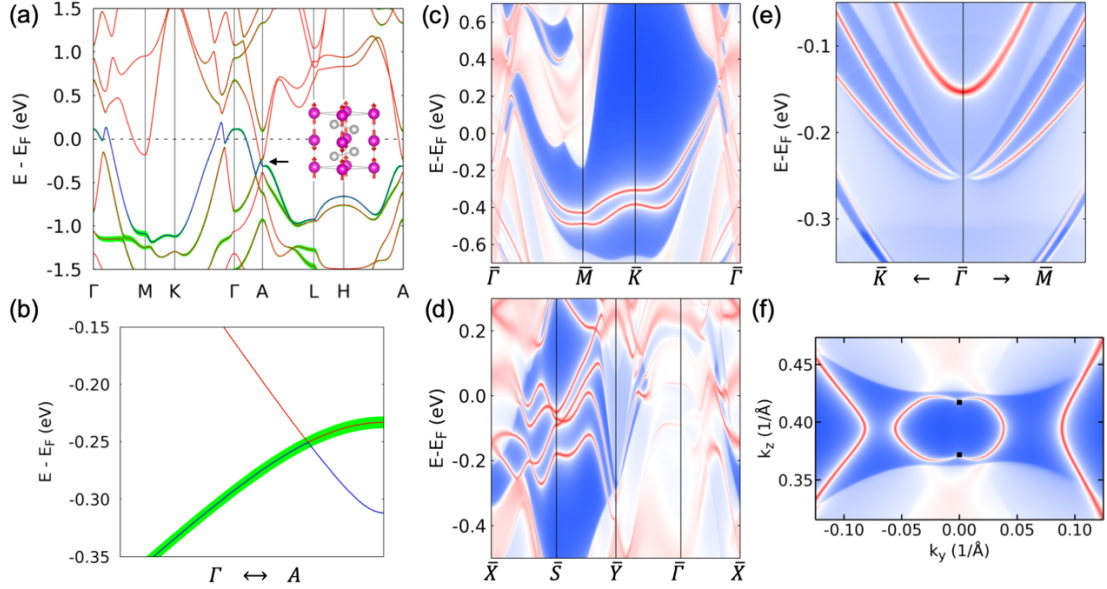

Figure 4. (a) Bulk band structure of EuTl$_2$ in PBE+SOC+U for A-type AFM with moment along *c*-axis (AFMA*c*). The inset shows the magnetic configuration. Black arrow points the bulk Dirac point and zoomed along *Γ-A* direction in (b). The valence band is in blue and the green shade stands for the projection of Tl $p_y$ orbitals. Panel (c) and (d) are the surface states on (001) and (010), respectively. (e) Surface states zoomed around $\bar{\Gamma}$ point on (001) (f) 2D Fermi surface (FS) on (100) at E$_F$–0.25 eV. The projections of the bulk Dirac points are labeled by black squares.

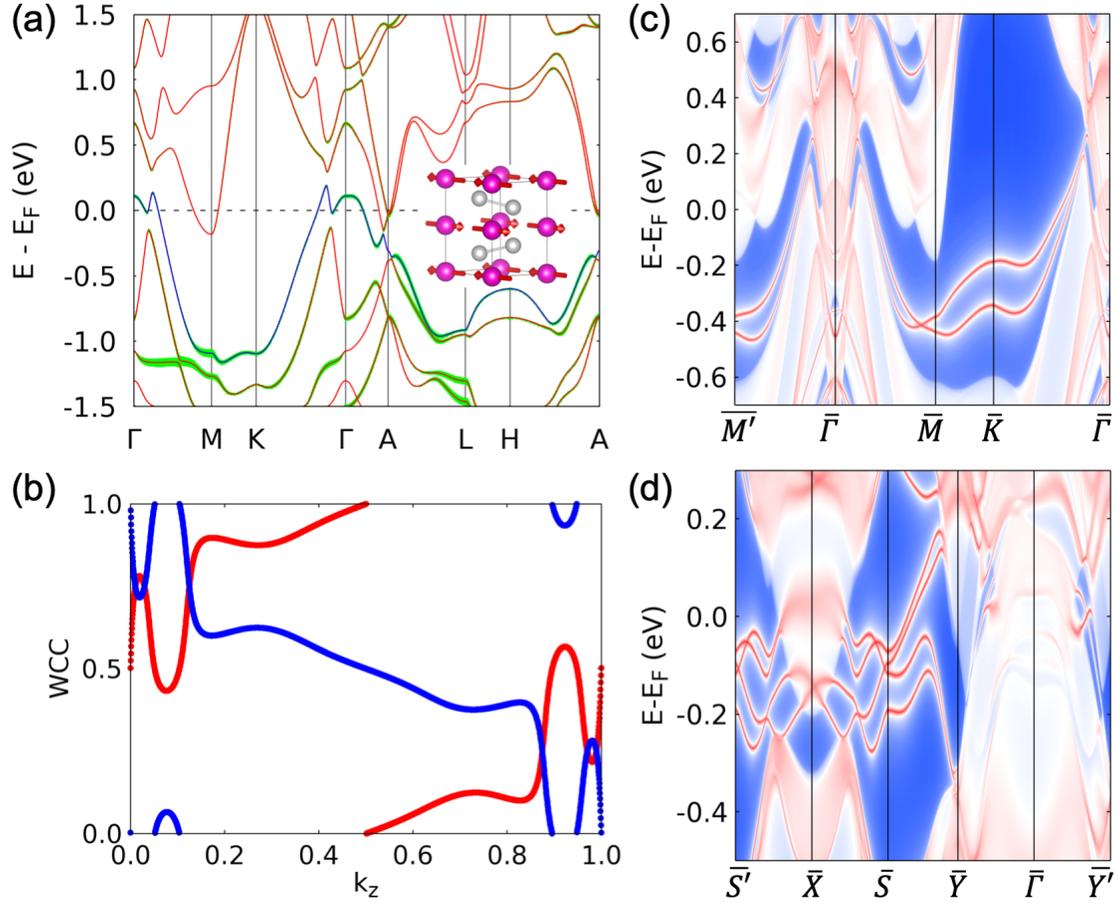

Figure 5. (a) Bulk band structure of EuTl$_2$ in PBE+SOC+U for A-type AFM with moment along *b*-axis (AFMA*b*). The inset shows the magnetic configuration. The valence band is in blue and the green shade stands for the projection of Tl $p_y$ orbitals. (b) WCC evolution of mirror eigenvalue of +i (red) and −i (blue) giving the $M_y$ MCN of 1. Panel (c) and (d) are the surface states on (001) and (010), respectively.

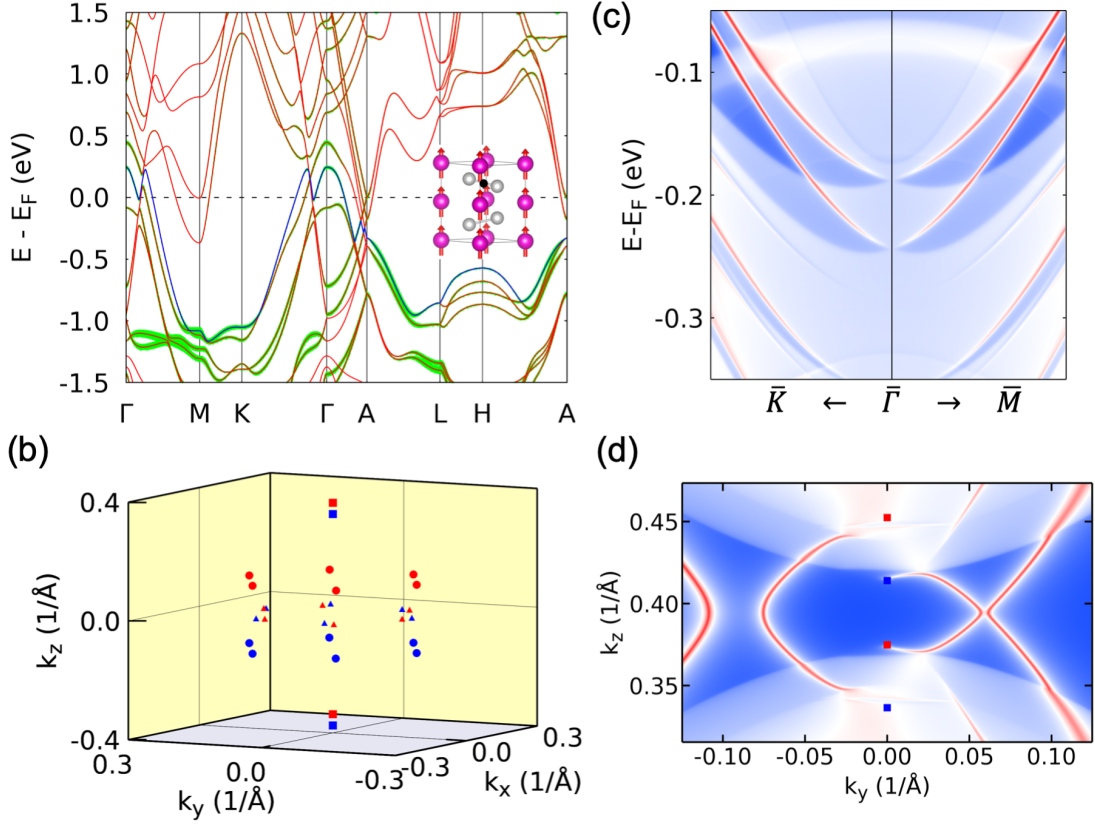

Figure 6. (a) Bulk band structure of EuTl$_2$ in PBE+SOC+U for FM with moment along $c$-axis (FM$c$). (b) Three sets of Weyl points (WPs) including two pairs along $k_z$ (squares) split from bulk Dirac points, six pairs parallelly separated across $k_z$=0 plane (circles) and six pairs on $k_z$=0 plane (triangles). Red (blue) stands for the chirality of +1 (−1). (c) The surface states on (001) around $\bar{\Gamma}$ point showing the two WPs split from the DP at $E_F$–0.25 eV and (d) the corresponding open Fermi arcs projected on the side (100) surface. The projections of the first set of WPs are labeled by red and blue squares.